\documentclass[]{interact}

\usepackage{epstopdf}
\usepackage{subfigure}
\usepackage{units}
\usepackage{braket}

\usepackage[numbers,sort&compress,merge]{natbib}
\bibpunct[, ]{[}{]}{,}{n}{,}{,}

\theoremstyle{plain}

\theoremstyle{definition}

\theoremstyle{remark}

\usepackage{color}
\usepackage{colortbl}

\begin{document}

\title{Phase- and intensity-dependence of ultrafast dynamics in hydrocarbon molecules in few-cycle laser fields}

\author{
\name{M. K\"ubel\textsuperscript{a}\thanks{Email: matthias.kuebel@physik.uni-muenchen.de. Present address: Joint Laboratory for Attosecond Science, National Research Council and University of Ottawa, Ottawa, Ontario K1A 0R6, Canada.},
C. Burger\textsuperscript{a,b},
R. Siemering\textsuperscript{c},
Nora G. Kling\textsuperscript{a},
B. Bergues\textsuperscript{a,b},
A.S. Alnaser\textsuperscript{d},
I. Ben-Itzhak\textsuperscript{e},
R. Moshammer\textsuperscript{f},
R. de Vivie-Riedle\textsuperscript{c},
and M.~F. Kling\textsuperscript{a,b}\thanks{Email: matthias.kling@lmu.de}
 }
\affil{
\textsuperscript{a}Department of Physics, Ludwig-Maximilians-Universit\"at, D-85748 Garching, Germany;
\textsuperscript{b}Max Planck Institute of Quantum Optics, D-85748 Garching, Germany;
\textsuperscript{c}Department of Chemistry and Biochemistry, Ludwig-Maximilians-Universit\"at, D-81377 Munich, Germany;
\textsuperscript{d}Physics Department, American University of Sharjah, POB26666, Sharjah, UAE;
\textsuperscript{e}J.R. Macdonald Laboratory, Physics Department, Kansas-State University, Manhattan, KS 66506, USA;
\textsuperscript{f}Max Planck Institute of Nuclear Physics, D-69117 Heidelberg, Germany.}
}
\maketitle

\begin{abstract}
In strong laser fields, sub-femtosecond control of chemical reactions with the carrier-envelope phase (CEP) becomes feasible. We have studied the control of reaction dynamics of acetylene and allene in intense few-cycle laser pulses at 750\,nm, where ionic fragments are recorded with a reaction microscope. We find that by varying the CEP and intensity of the laser pulses it is possible to steer the motion of protons in the molecular dications, enabling control over deprotonation and isomerization reactions. The experimental results are compared to predictions from a quantum dynamical model, where the control is based on the manipulation of the phases of a vibrational wave packet by the laser waveform. The measured intensity dependence in the CEP-controlled deprotonation of acetylene is well captured by the model. In the case of the isomerization of acetylene, however, we find differences in the intensity dependence between experiment and theory. For the isomerization of allene, an inversion of the CEP-dependent asymmetry is observed when the intensity is varied, which we discuss in light of the quantum dynamical model. The inversion of the asymmetry is found to be consistent with a transition from non-sequential to sequential double ionization.
\end{abstract}


\section{Introduction}
In the past few decades, strong-field lasers have proven to be valuable tools for the manipulation of chemical bonds. In order to achieve and observe such phenomena, experimental and theoretical efforts, in particular on small molecules, have been extensive (see e.g. \cite{shapiro:12}). Understanding dynamic systems on a small scale is a prerequisite for moving towards more complex molecules and reactions, and ultimately their control. Molecular processes and chemical reactions are governed by nuclear motion and the motion of the valence electrons. Control of the reaction to give desired products is therefore achieved through the manipulation of both nuclear and electronic motion, on their respective time scales. It depends on the molecular system, whether it is the nuclear or electron dynamics, or both, that offer the most efficient control knob. One possible approach to address these motions is to illuminate molecules with intense few-cycle light pulses. Tailoring the electric-field waveform of optical pulses on sub-cycle timescales \cite{baltuska:03,goulielmakis:07,Krausz2009,cerullo2011} opens the door to the control of electron dynamics in atoms and molecules on their natural motion timescales. A suitable parameter to modify the electric-field waveform of a few-cycle pulse, $E(t) = E_0(t) cos(\omega t + \phi)$, with envelope $E_0(t)$, and carrier frequency $\omega$, is the carrier-envelope phase (CEP) $\phi$. The CEP-control of molecular dynamics has been predominantly investigated for diatomic molecules in both experiment and theory \cite{Kling2013,lepine:13}. A large part of previous work was performed on molecular hydrogen and molecular hydrogen ions (see e.g.~ \cite{bandrauk:04,ben-itzhak:04,kling:06,geppert:08,kremer:09,Kelkensberg2011,Anis2012,Xu2013,Kling2013b,Rathje2013,Li2014}).

Even in these simple systems, the strongly driven, coupled (and correlated) electron-nuclear dynamics typically need to be described by models beyond the Born-Oppenheimer approximation. For a theoretical treatment of molecular dynamics induced by intense few-cycle pulses, the nuclear and electron dynamics need to be included in a coupled manner. However, a theoretical description of electron dynamics in multi-electron molecules is a challenge, and its appropriate treatment is the aim of state-of-the-art research. Many approaches use time-dependent analogs of well-established quantum chemical methods. Based upon the time-dependent Hartree-Fock theory \cite{kulander:87} and the (explicitly) time-dependent density-functional theory \cite{Runge1984} there are many expansions to incorporate electron correlation and make use of post-Hartree-Fock methods like time-dependent-configuration-interaction (TD-CI) \cite{klamroth:03,Rohringer2006} time-dependent multi-configuration-self-consistent-field (TD-MCSCF) \cite{kato:04} and multi-configuration time-dependent-Hartree-Fock (MC-TDHF) \cite{zanghellini:04}. In other approaches the electronic wavefunction is directly propagated using Green's function formalism \cite{kuleff:05} or on the basis of molecular orbitals \cite{remacle:06}. When molecular reactions are considered, the nuclear motion needs to be included, preferably on equal footing. For the three-body-system D$_2^+$ the coupled dynamics can be fully calculated quantum mechanically \cite{bandrauk:04,roudnev:07}. A method that includes the valence electrons \cite{geppert:08} and the nuclear dynamics on the quantum level has been
successfully applied to  larger diatomics like CO \cite{Kling2013} and K$_2$ \cite{Bayer2013}. Moreover, a multi-configuration electron-nuclear dynamics method \cite{nest:09} exists, which may handle more than two nuclei and one electron. 	
While these methods describe the electronic and nuclear motion very accurately, the calculations are computationally demanding and have not yet been realized for larger molecular systems. 

Experimental carrier-envelope phase control studies on polyatomic molecules and complex reactions are only recently emerging (see e.g. \cite{Xie2012,mathur:13,Alnaser2014,Miura2014,Kubel2016_prl,Li2016}). CEP effects in the deprotonation of acetylene, induced by few-cycle laser pulses, have attracted considerable interest. In particular, a strong CEP dependence in the total fragmentation yield has been reported \cite{Xie2012}. Moreover, it has been demonstrated that controlling the CEP permits selective breaking of either C-H bond, leading to directional proton emission \cite{Alnaser2014,Miura2014}. Very recently, we have reported on the control of the preferential direction of hydrogen migration in acetylene and allene \cite{Kubel2016_prl}, and toluene \cite{Li2016} using the CEP of a few-cycle laser pulse. Here, we go beyond studies at a single intensity of the laser field and report on combined CEP and intensity studies of the strong-field-induced dissociative ionization of acetylene and allene.

\section{Methods}
\subsection{Experimental methods}\label{experiment}

Intense few-cycle laser pulses are obtained by focusing the output of the SMILE laser, which is based on a commercial Femtolasers Femtopower HR system, into a gas-filled hollow-core fiber. Depending on the input pulse energy, the fiber is filled with $\approx 0.7$\,bar of Ar for pulse energies below 0.5\,mJ, or $\approx 3.0$\,bar of Ne for powers above. In either case, a broadband supercontinuum supporting 4-fs pulses at 750\,nm central wavelength is obtained. The pulses are compressed using chirped multilayer mirrors \cite{Pervak2009} and fused silica wedges. The pulse duration is verified using a home-built transient-grating frequency-resolved optical gating setup \cite{Kane1993, Schultze2011}. In the SMILE laser the CEP of the laser oscillator is stabilized using the feed-forward technique \cite{Lucking2012}. The CEP of the amplified laser pulses is measured using a stereo-ATI phase meter \cite{Rathje2012} (in the acetylene experiments) or an f-2f interferometer \cite{Schultze2011} (in the allene experiments) after the hollow-core fiber. To control the CEP, the dispersion within the stretcher unit of the laser amplifier is varied.

The few-cycle laser pulses are focused ($ f=\unit[17.5]{cm}$) into a cold gas jet of neutral hydrocarbon molecules in the center of a reaction microscope (REMI) \cite{Ullrich2003}. The base pressure in the REMI is kept below $\unit[10^{-10}]{mbar}$ to minimize background signals. The number of molecules in the laser focus is controlled by changing the thickness of the gas target along the laser propagation direction using a slit with adjustable width that cuts into the gas stream. Ions generated in the laser focus are directed onto a time- and position-sensitive multi-hit capable detector by means of a homogeneous electric field ($\approx \unit[30]{Vcm^{\scriptscriptstyle -1}}$). The measurement of the impact time and positions of ions on the detector provides the three-dimensional (3D) momentum distributions. Molecular break-up channels are identified using photoion-photoion coincidence (so called PIPICO) time-of-flight spectra. By testing the momentum sum of coincident ions, i.e. $p_i = p_{A,i} + p_{B,i}$ for momentum conservation, fragments originating from the same parent molecule are selected. Here, $i=\{x,y,z\}$ denote the dimension in the laboratory frame, and $A$ and $B$ denote two ionic fragments. We note that the center-of-mass momentum $p_i$ is not exactly zero, but equals the negative momentum sum of all emitted electrons. Therefore, the center-of-mass momentum distribution provides information on the ionization process preceding the molecular break-up.

\begin{figure}
\centering
\includegraphics[width=\textwidth]{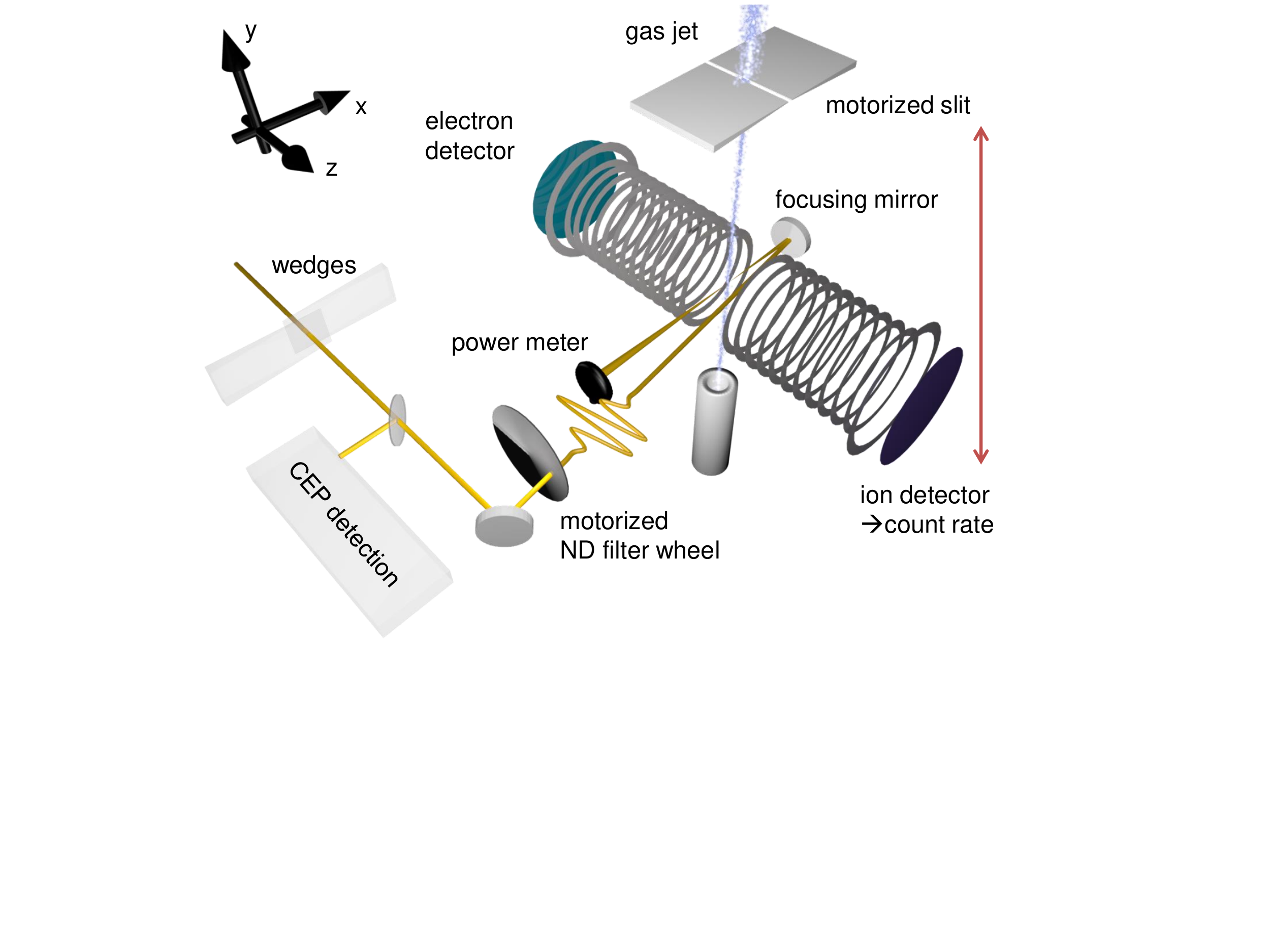}
\vspace{-4cm}
\caption{Schematic of the simultaneous CEP and intensity-scanning and -tagging technique employed in the experiments. See text for details. The laser field is linearly polarized along the $z$-axis. The red arrow indicates the feedback from the ion count rate to the motorized slit.} \label{fig:experiment}
\end{figure}

In order to facilitate the coincident detection of ions, the average count rate on the ion detector is limited to less than one event per laser shot. Maintaining this condition while varying the laser intensity is challenging, due to the highly non-linear intensity dependence of strong-field ionization. Here, we employ an intensity scanning method sketched in Figure \ref{fig:experiment}. The laser pulse energy sent into the REMI is adjusted with a motorized neutral density (ND) filter wheel. Based on the ion count rate, the data acquisition computer generates a feedback signal to control the number of molecules available for ionization using the adjustable slit. Thus, a moderate count rate is maintained while the intensity is scanned over a typical range from $\approx \unit[1\times 10^{14}]{W/cm^2}$ to $\approx \unit[3\times 10^{14}]{W/cm^2}$. At the exit of the REMI, the laser power is measured with a fast electronic power meter. Since pulse duration and focusing geometry of the laser pulses remain unaffected by the ND filter setting, the measured laser power is directly proportional to the intensity in the laser focus. The proportionality factors between the laser intensity and pulse energy are determined in separate measurements on Ar and Ne targets \cite{Kubel2016_pra}. The data recorded with the REMI are correlated with both the simultaneously measured CEP and power of the laser pulses in order to access the CEP and intensity-dependence of the deprotonation, 
\begin{align*}
\mathrm{C_2H_2^{2+}} &\rightarrow \mathrm{H^+} + \mathrm{C_2H^+}, \\
\mathrm{C_3H_4^{2+}} &\rightarrow \mathrm{H^+} + \mathrm{C_3H_3^+}
\end{align*}
 and isomerization, 
 \begin{align*} 
 \mathrm{C_2H_2^{2+}} &\rightarrow \mathrm{C^+} + \mathrm{CH_2^+}, \\
 \mathrm{C_3H_4^{2+}} &\rightarrow \mathrm{H_3^+} + \mathrm{C_3H^+},
 \end{align*}  
  channels.

\subsection{Theoretical methods}
Experimental results are discussed within the framework of a quantum dynamical model, where the CEP dependence of photochemical reactions arises from the preparation and manipulation of a multimode vibrational wavepacket. The model has been employed previously to explain the CEP dependence of the deprotonation of acetylene \cite{Alnaser2014}, the isomerization of acetylene, and the isomerization of allene \cite{Kubel2016_prl}. The details for the treatment of deprotonation and isomerization reactions can be found in Refs \cite{Alnaser2014, Kubel2016_prl}. Here, we provide a general description of the control mechanism, focusing on the commonalities of the two reactions.

The simulations are performed in two steps: In a first step, a nuclear wavepacket is prepared by the interaction of the initially neutral molecules with a few-cycle laser field. After the first step, the laser pulse has passed and the molecule is left in a reactive state, which, for our conditions, is an excited state of the molecular dication, from which deprotonation or isomerization occurs. In the second step, the prepared nuclear wavepacket is propagated on the potential energy surface (PES) of the reactive state. From the results of the wave packet propagation, the CEP dependence of the reaction is evaluated.

\begin{figure}
\centering
\includegraphics[width=\textwidth]{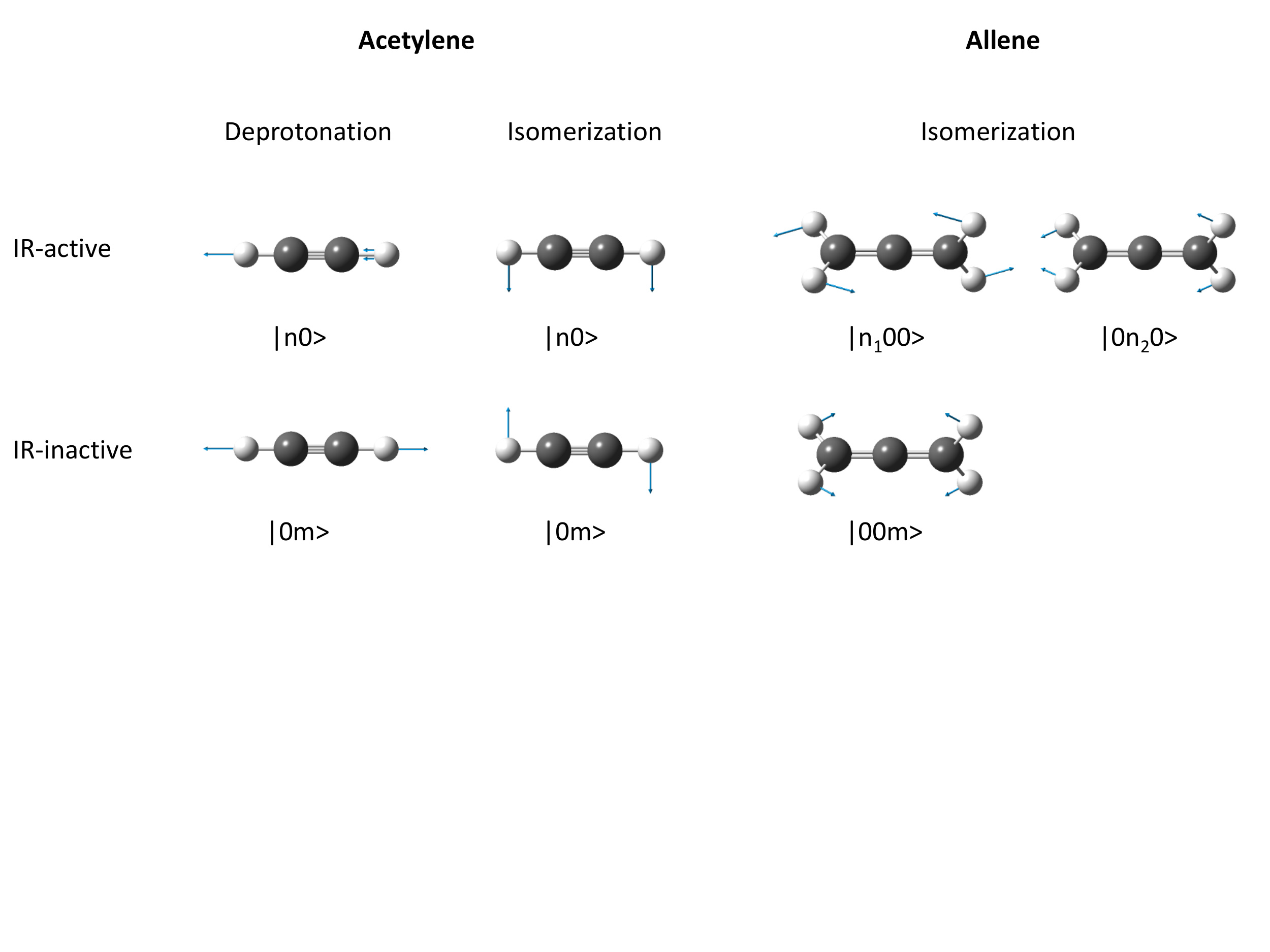}
\vspace{-4.5cm}
\caption{Illustration of the relevant IR-active and IR-inactive normal modes during the initiation of the deprotonation and isomerization of acetylene, and of the isomerization of allene. For each mode, the notation used in the text, is given.} \label{fig:theory}
\end{figure}

In the first simulation step, the nuclear dynamics are described in the basis of normal modes. To keep the computational effort tolerable, the most important normal modes for the initiation of a reaction are identified. In Figure \ref{fig:theory} the relevant modes for the initiation of each reaction are displayed. Out of these modes, only IR-active modes can be excited by the laser field. In the following, we denote the IR-active modes for acetylene as $\ket{n0}$ and the IR-inactive modes as $\ket{0m}$, where $n, m$ is the number of vibrational quanta, respectively. Here, the respective time evolution factor $\exp{\left(-i \frac{E_{m/n}}{\hbar} t\right)}$ is implicitly included and the relative phase of the eigenfunctions is set to compensate any phase offset relative to the CEP. In the case of allene, where three modes are used, the notation is $\ket{n_100}$, $\ket{0n_20}$, and $\ket{00m}$, respectively. 

The interaction of the neutral molecules with the external laser field is calculated on two-dimensional PESs (or three-dimensional in the case of allene) along the aforementioned vibrational modes by solving the time-dependent Schr\"odinger equation (TDSE)
\begin{equation}
i \hbar \frac{\partial d}{\partial t} \Psi_n (t) = \left(H_n + \mu_{nn} \epsilon \left( t \right) \right) \Psi_n \left( t \right),
\end{equation}
where $H_n$ is the Hamiltonian of the ground state, $\mu_{nn}$ is the associated dipole moment and, $\Psi_n \left( t \right)$ is the nuclear wave function. The light-field $\epsilon(t)$ is included in the dipole approximation and is characterized by a full width at half maximum of the intensity envelope of 4\,fs, and a carrier wavelength of 750\,nm.

The interaction of the molecule with the off-resonant laser field leads to a transient CEP-dependent population in the IR-active vibrational modes $\ket{n0} e^{-i\phi}$ (or $\left(\ket{n_100}+\ket{0n_20}\right) e^{-i\phi}$ in the case of allene). If the molecule remains in the electronic ground state, the vibrational excitation vanishes at the end of the pulse. If the molecule is ionized, however, the population of the IR-active vibrational modes can persist until after the laser pulse has passed. Moreover, due to the projection of the vibrational wave packet on the cationic PES, also IR-inactive modes are populated, independent of the CEP. The CEP-dependence of the superposition of vibrational modes is essentially contained in the following basic wave packets
\begin{align}
\Psi_\mathrm{basic} &= \ket{10} e^{-i\phi} +\ket{01} &\mathrm{for~acetylene,~or} \\
\Psi_\mathrm{basic} &=\left(\ket{100}+\ket{010}\right) e^{-i\phi} +\ket{001} &\mathrm{for~allene.}
\end{align}

In the simulations, it is assumed that the molecule is ionized at $t=T$, i.e. when the laser field reaches its intensity maximum. A second ionization step is assumed to occur via electron recollision at $t=T + 0.75 \mathrm{oc}$, where $\mathrm{oc}=\unit[2.5]{fs}$ is an optical cycle. The second ionization transfers the molecule to a reactive state, which is the lowest excited state of the molecular dication that supports deprotonation or isomerization. The reactive states are the $A^3\Pi$ state for acetylene, and the $B^3\Pi$ state in the case of allene.

In the second simulation step, the prepared wave packets are propagated on the PES of the reactive state. The wave packet propagation has been discussed in detail in Ref.~\cite{Alnaser2014} for the deprotonation of acetylene and in Ref.~\cite{Kubel2016_prl} for the isomerization of acetylene. Briefly, the wave packet possesses a preferential propagation direction, which is determined by the CEP. Depending on this direction, the wave packet preferentially propagates along certain paths that correspond either to left or right deprotonation or isomerization, with respect to the laser polarization. The reaction yields for the left and right deprotonation or isomerization, $L$ and $R$, respectively, are evaluated as a function of CEP from the results of the wave packet propagation. Finally, the asymmetry parameter,
\begin{equation}
A(\phi)=\left(R(\phi)-L(\phi)\right) / \left(R(\phi)+L(\phi)\right).
\end{equation}
is calculated and used to quantify the CEP control for each reaction.

In essence, the CEP dependence of the reactions rely on two aspects. First, a CEP-dependent population of IR-active vibrational modes is transiently excited by the laser field. Second, the molecule is ionized at a single instance during the laser pulse, which leads to the coherent excitation of IR-inactive modes. The resulting wavepacket depends on the CEP of the laser through the phase of the IR-active modes.
The CEP-dependence of the wavepacket results in a preferential propagation direction of the wavepacket. When propagated on a reactive state, the propagation direction translates into a preference in the deprotonation or isomerization direction, which is steerable by the CEP.

\section{Results and Discussion}

\subsection{Double ionization of acetylene}

Both reactions, deprotonation and isomerization, occur in the dication. Therefore, they are preceded by double ionization of the neutral molecules by the intense few-cycle laser field. Characterizing the ionization process provides valuable information on the field-driven molecular dynamics. Very recently, it has been demonstrated for argon that the combined intensity and CEP dependence of double ionization allows characterizing the underlying mechanisms \cite{Kubel2016_pra}. Here, we extend this technique to molecular double ionization.

\begin{figure}
\centering
\includegraphics[width=\textwidth]{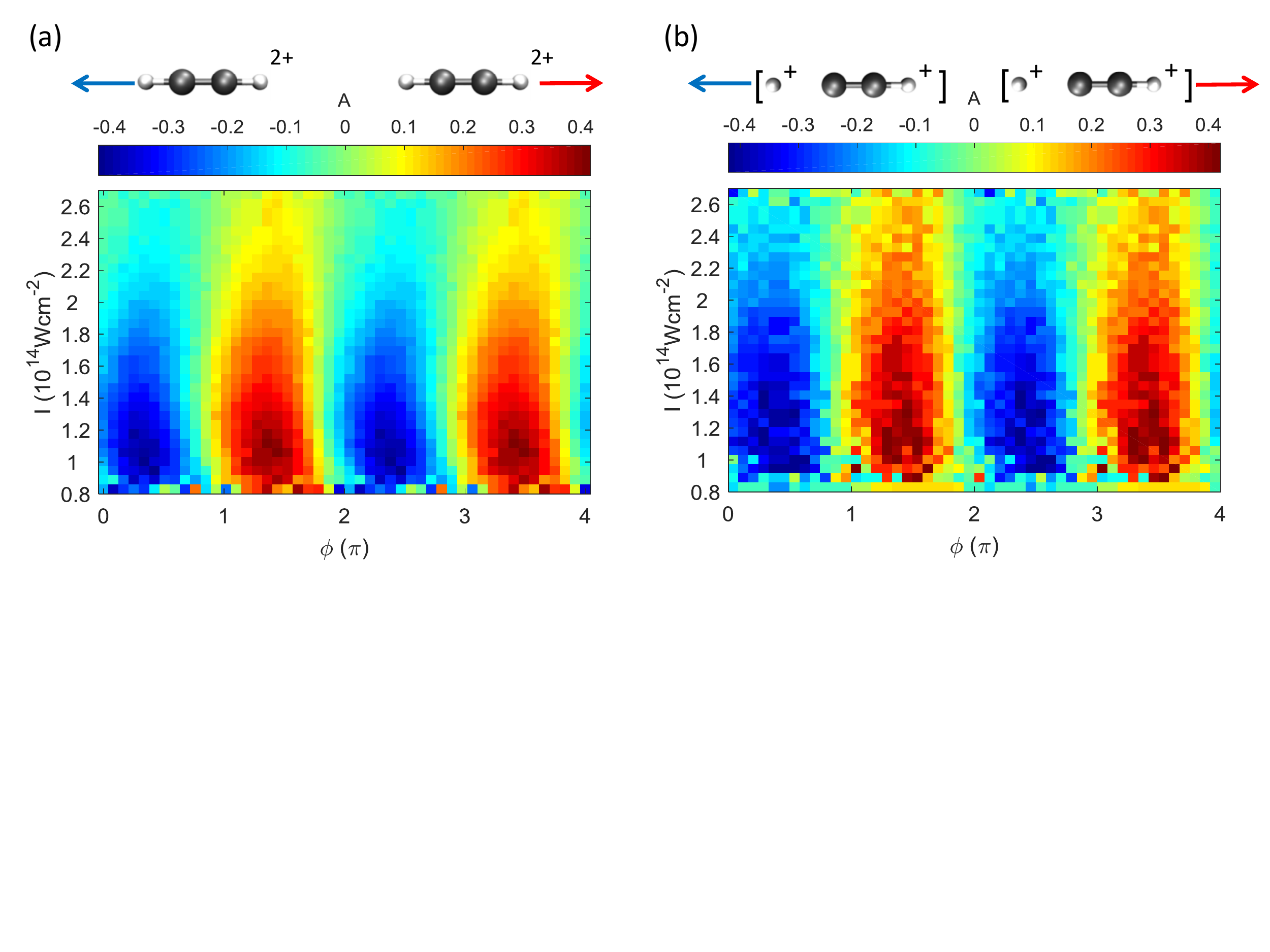}
\vspace{-4.5cm}
\caption{Asymmetry in the recoil momentum distributions for double ionization of acetylene, recorded as a function of CEP and laser intensity. Panel (a) is for acetylene dications (C$_2$H$_2^{2+}$), panel (b) is for the center-of-mass motion of the deprotonation fragments (H$^+$ + C$_2$H$^{+}$), see text for details. The cartoons on top of the diagrams indicate the ionization channel. } \label{fig:double_ionization_acetylene}
\end{figure}

In Figure \ref{fig:double_ionization_acetylene}a, the combined intensity and CEP dependence of the asymmetry in the C$_2$H$_2^{2+}$ recoil momentum spectra is shown. The data exhibits the usual $2\pi$ periodicity in CEP \cite{roudnev:07}. The strong modulation amplitude at low intensity decays when the intensity is increased. The phase offset of the CEP-dependent asymmetry (in brief, the asymmetry phase, in other works referred to as the ``phase of the phase" \cite{Bauer2015}) exhibits only a slight intensity dependence.

The intensity-dependent behavior of the C$_2$H$_2^{2+}$ asymmetry is similar to the one reported for Ar$^{2+}$ in Ref.~\cite{Kubel2016_pra}. There, a large asymmetry amplitude at low intensity, that decays towards higher intensities, as well as a relatively flat intensity dependence of the asymmetry phase, were identified as typical behavior of recollision-induced, non-sequential double ionization (NSDI). Hence, NSDI represents the dominant mechanism for the production of C$_2$H$_2^{2+}$ in the entire intensity range studied in the present experiments.

Having characterized the double-ionization process, we turn towards molecular reactions following double ionization. In a first step, the center-of-mass motion of the deprotonation fragments is analyzed in order to obtain information on the ionization process preceding the molecular break-up. The intensity and CEP-dependence of the asymmetry of the center-of-mass motion of the deprotonation fragments is shown in Figure \ref{fig:double_ionization_acetylene}b. The signal behaves similar to the one shown in Figure \ref{fig:double_ionization_acetylene}a, indicating that in our experiment the deprotonation of acetylene is also induced by electron recollision.

We note that the sligthly larger asymmetry amplitudes in the dication momentum distributions, measured for the deprotonation channel (Figure \ref{fig:double_ionization_acetylene}b) as compared to the non-dissociative channel (Figure \ref{fig:double_ionization_acetylene}a), are consistent with the electron recollision scenario. Deprotonation occurs from excited states of the dication \cite{Doblhoff-Dier2016}. Therefore, populating the relevant states for deprotonation may require a higher recollision energy than populating the electronic ground state. The requirement of higher recollision energy has been shown to lead to an increased asymmetry amplitude for molecular double ionization \cite{Kubel2014}.

\subsection{Deprotonation of acetylene}
For the deprotonation of acetylene, a very strong CEP dependence of the total deprotonation yield has been reported to occur at the onset of recollision-driven double ionization \cite{Xie2012}. Our dedicated intensity scanning technique should be well suited to shed more light on this process. However, we measure the CEP-induced modulation in the total deprotonation yield to be smaller than 5\% for any intensity in our experiment.

\begin{figure}
\centering
\includegraphics[width=\textwidth]{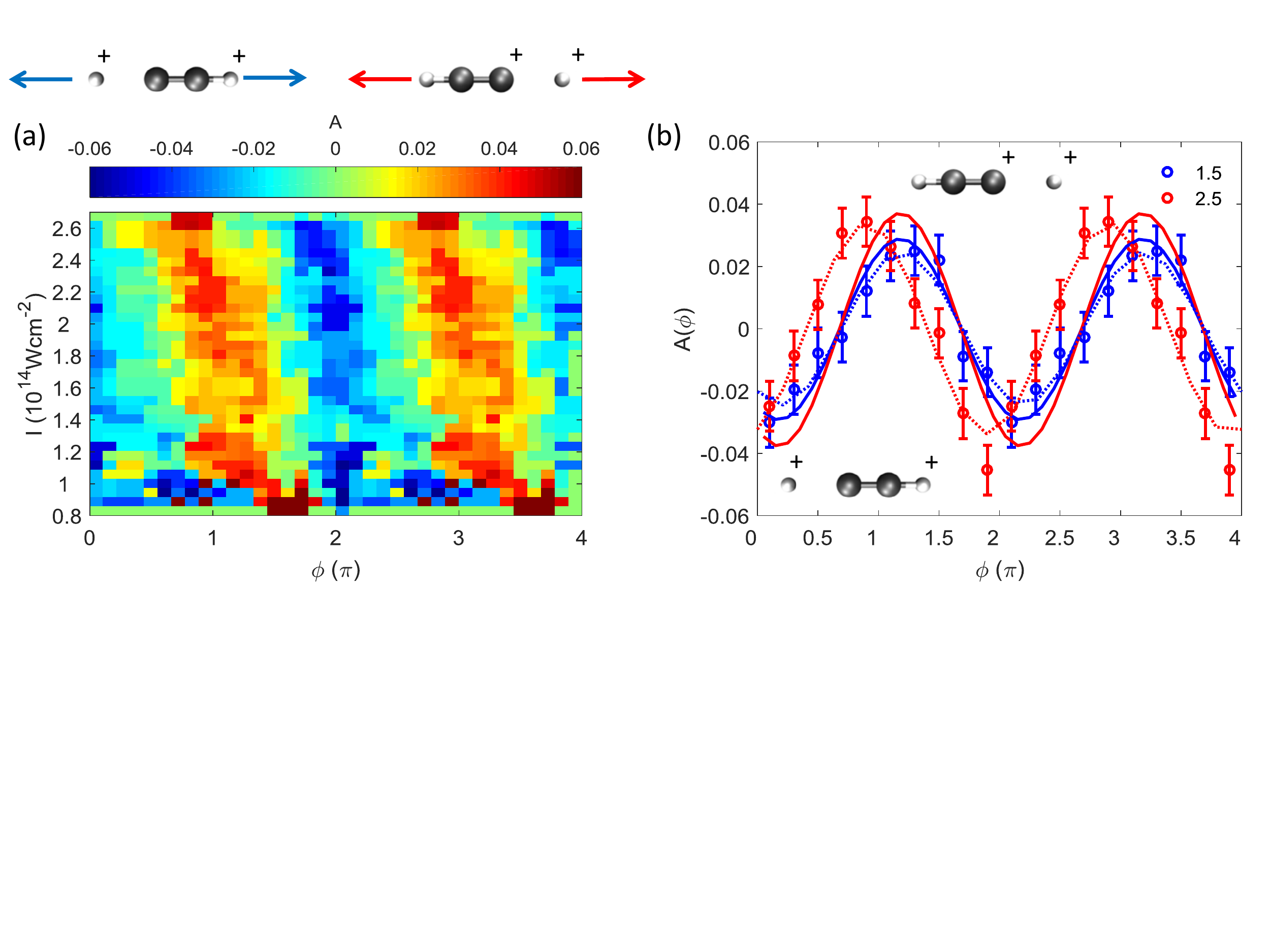}
\vspace{-4cm}
\caption{(a) Asymmetry in the proton momentum distributions for the deprotonation of acetylene, recorded as a function of CEP and laser intensity. (b) Comparison of the recorded CEP-dependent deprotonation asymmetry (symbols) to the calculated one (solid lines) for two different intensities ($\unit[1.5 \times 10^{14}]{W/cm^2}$, blue, and $\unit[2.5 \times 10^{14}]{W/cm^2}$, red). The experimental data was averaged over a range of $\pm \unit[0.2 \times 10^{14}]{W/cm^2}$. The CEP axis for the experiment was shifted for best agreement with the theory at $I = \unit[1.5 \times 10^{14}]{W/cm^2}$. The dotted lines represent sinusoidal fits of the experimental data.} \label{fig:deprotonation_acetylene}
\end{figure}

Instead of the CEP-dependence of the total yield, we focus on the directional yields and show the combined intensity and CEP dependence of the asymmetry in the momentum distribution of protons, emitted from acetylene, in Figure \ref{fig:deprotonation_acetylene}a. The CEP dependence of the deprotonation is robust with respect to intensity variation, with a slight trend of increasing asymmetry amplitude with increasing intensity. The asymmetry phase exhibits a small drift towards smaller values with increasing intensity.

A direct comparison of the measured data to theory is given in Figure \ref{fig:deprotonation_acetylene}b. The predicted asymmetry amplitudes agree well for both intensities. However, the phase shift of $(50 \pm 13)^\circ$, observed in the experimental data, is not reproduced by theory. The phase shift probably originates from an alteration in the underlying ionization dynamics with increasing intensity, given the observed change in the CEP-dependent asymmetry in the double ionization, shown in Figure \ref{fig:double_ionization_acetylene}. In the model, the ionization dynamics are not altered when the intensity is increased. Instead the simplifying assumption is made that the molecule is only ionized at the intensity maximum of the laser pulse. This assumption may lead to an overestimation of the asymmetry amplitude \cite{Kubel2016_prl}, which is, however, not observed. The reason for the relatively large asymmetry observed in the experiment, may lie in the contributions from higher excited states, which is discussed in the following section.

\subsubsection{Contributions from different electronic states}

\begin{figure}
\centering
\includegraphics[width=\textwidth]{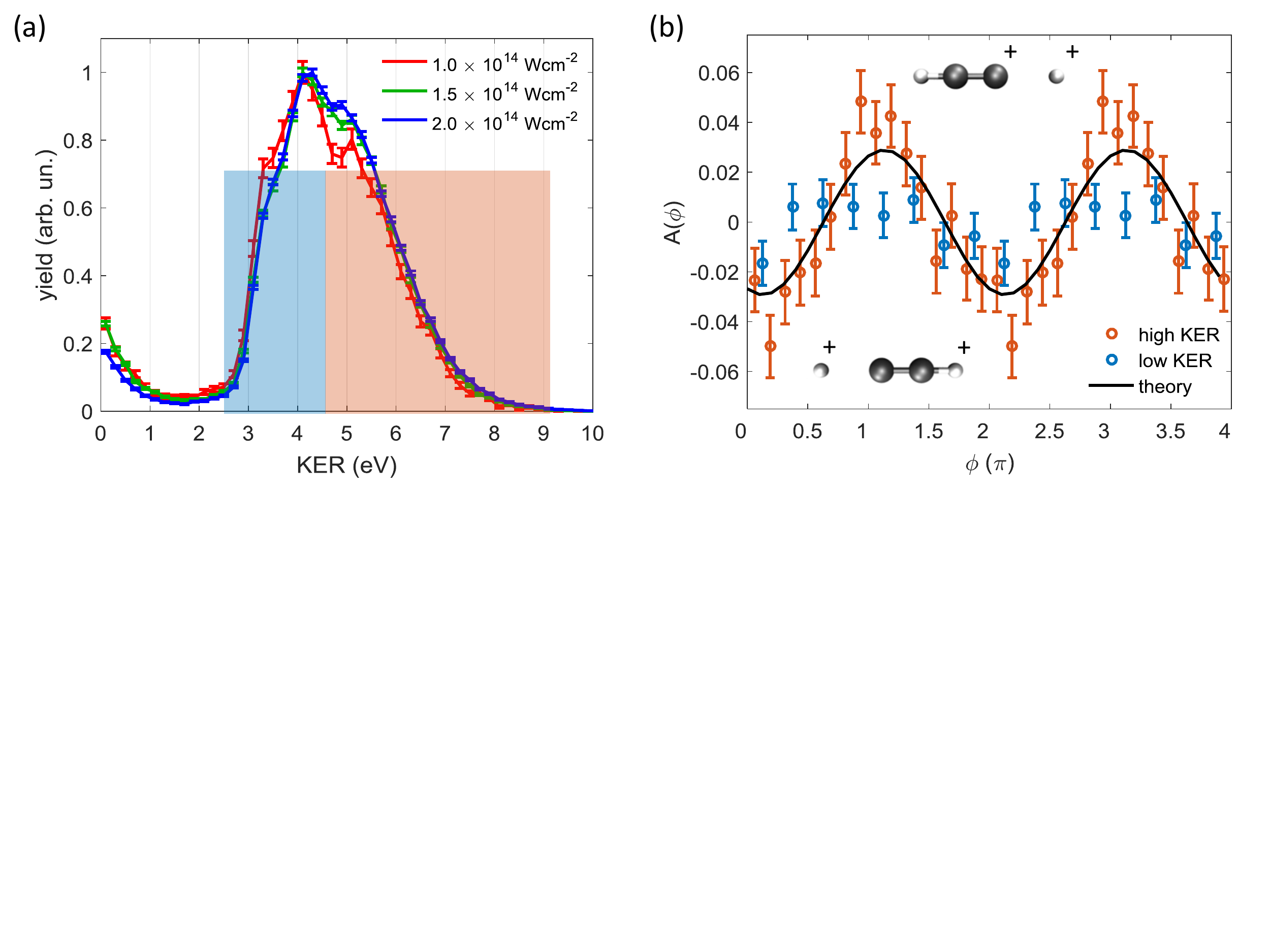}
\vspace{-5.5cm}
\caption{(a) KER distribution measured for the deprotonation of acetylene, for three different intensity values, each integrated over $\pm \unit[0.2 \times 10^{14}]{W/cm^2}$. The blue and red shaded areas mark the low- and high-KER channels, respectively. (b) CEP-dependent asymmetries in the proton momentum distributions for the low (blue) and high KER (red) channels at an intensity $I = \unit[1.5 \times 10^{14}]{W/cm^2}$. The solid black line depicts the theoretical result for the same intensity, assuming the $A^3\Pi$ state of the dication for the reaction.}

\label{fig:deprotonation_channels_acetylene}
\end{figure}

In Figure \ref{fig:deprotonation_channels_acetylene}a, the kinetic energy release (KER) distribution for the deprotonation of acetylene is displayed for three different intensity values. The KER spectrum consists of a broad distribution spanning from approximately 2.5 to 8\,eV, with a pronounced shoulder near 5\,eV that is enhanced at high intensities. This suggests that the high KER values originate from one or more higher excited states than the low KER values. Therefore, the onset of the shoulder is used to separate the deprotonation signal into low-KER and high-KER channels.

In Figure \ref{fig:deprotonation_channels_acetylene}b it is shown that the high-KER channel exhibits a much stronger CEP-dependent asymmetry than the low-KER channel. The comparison to the theoretical result shows that the measured asymmetry in the high-KER channel is even larger than the one predicted by the simulations. For the low-KER channel, however, the asymmetry is overestimated by theory. Note that, for computational reasons, the simulations use only a single reactive state, which is the lowest excited dicationic state that leads to deprotonation. The overestimation of the asymmetry amplitude for low KER values by theory may originate from the simplifying assumption that ionization occurs only at the peak of the laser field.

\subsection{Isomerization of acetylene}

The isomerization channel, leading to dissociation of acetylene into C$^+$ + CH$_2^+$, exhibits a narrow KER distribution around 4.4\,eV, which agrees with previous studies using strong IR laser pulses \cite{Osipov2003, Alnaser2006}. Within the KER distribution there are no discernible contributions from different states with different CEP dependencies.

\begin{figure}
\centering
\includegraphics[width=\textwidth]{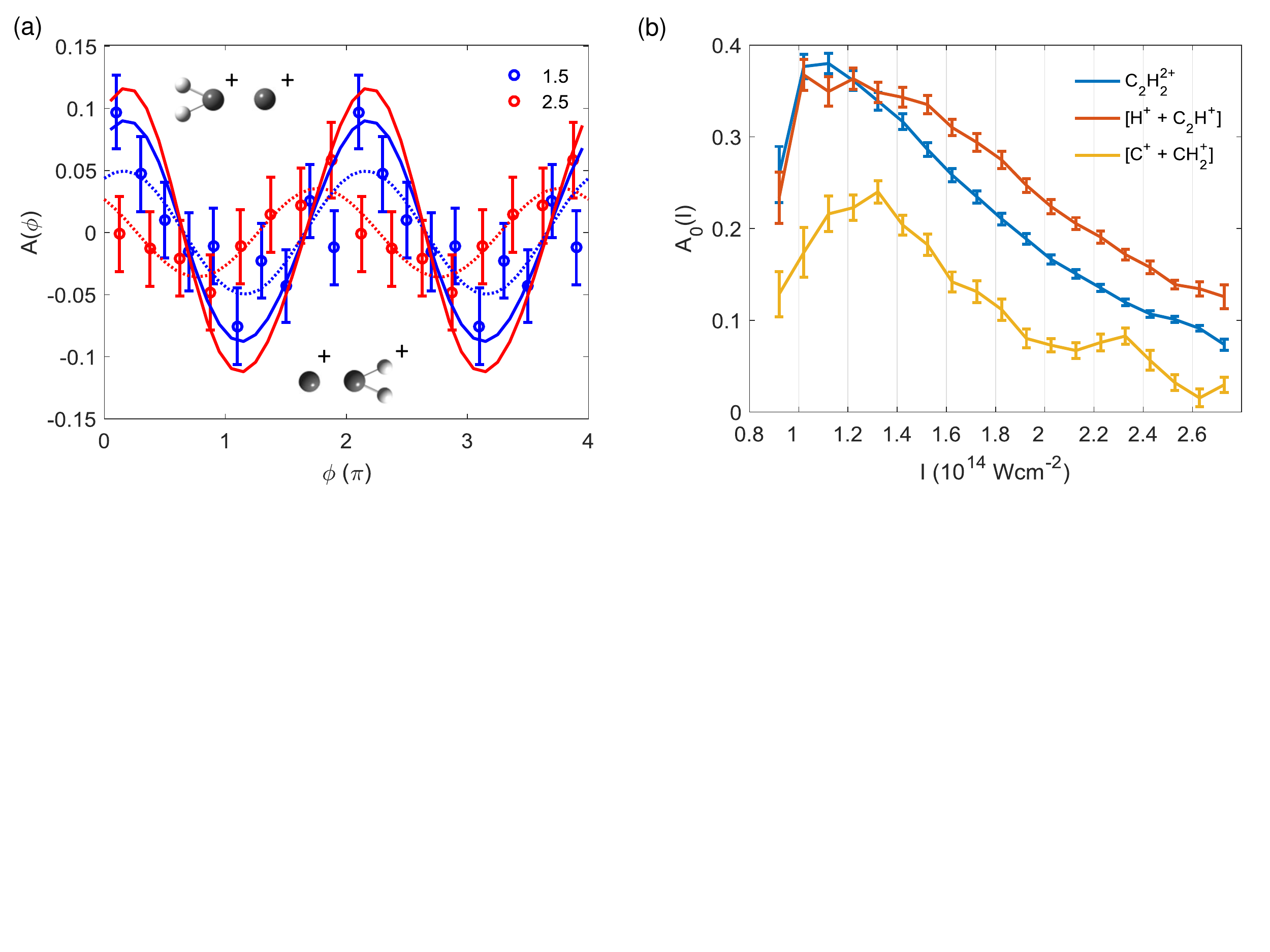}
\vspace{-5.5cm}

\caption{(a) CEP-dependent asymmetries in the momentum distributions of carbon ions emitted from vinylidene after hydrogen migration in acetylene, for two different intensities, $\unit[1.5 \times 10^{14}]{W/cm^2}$ (blue) and $\unit[2.5 \times 10^{14}]{W/cm^2}$ (red). The data is averaged over a range of $\pm \unit[0.3 \times 10^{14}]{W/cm^2}$ for each intensity value. The dotted lines are sinusoidal fits of the experimental data (symbols). The solid lines represent the calculated asymmetry curves. (b) Measured intensity dependence of the asymmetry amplitudes of the center-of-mass motion for the non-dissociative (blue), deprotonation (red) and isomerization (orange) channels.}

\label{fig:isomerization_acetylene}
\end{figure}

In Figure \ref{fig:isomerization_acetylene}a, the asymmetry in the momentum distribution of C$^+$ ions, detected in coincidence with CH$_2^+$ is displayed for two intensity values. For the high intensity, a significantly smaller asymmetry is observed than for low intensity. The asymmetry at high intensity exhibits a phase shift of $(70 \pm 20)^\circ$ with respect to the asymmetry at low intensity. The magnitude and direction of the observed phase shift is consistent with the one observed for the deprotonation channel, suggesting that it also originates from the underlying ionization dynamics.
The decrease in the asymmetry amplitude, however, contrasts the observations made for the deprotonation channel. It is also opposite to the theoretical prediction that the asymmetry amplitude should increase with increasing intensity. 
Generally, the calculations tend to overestimate the asymmetry amplitudes significantly more than in the case of the deprotonation of acetylene. We attribute this overestimation to the rotation of the acetylene molecule, occuring as a consequence of the hydrogen migration \cite{Osipov2003}. This rotation is not accounted for in the theory.

To gain more information on the initiation of the hydrogen migration, we analyze the asymmetry of the center-of-mass motion of the isomerization fragments and compare it to the deprotonation and non-dissociative double-ionization channels. The intensity-dependence of the center-of-mass asymmetry amplitude for all three channels is displayed in Figure \ref{fig:isomerization_acetylene}b. While the deprotonation channel exhibits a stronger center-of-mass asymmetry than the non-dissociative channel for most intensities, the asymmetry in the isomerization channel is weaker. We recall that the center-of-mass asymmetry is solely determined by the momenta of the photoelectrons, which are emitted before the hydrogen migration takes place. Hence, the molecular rotation, induced by the hydrogen migration, cannot affect the center-of-mass asymmetry. Instead, the different asymmetry amplitudes indicate a difference in the ionization processes preceding isomerization and deprotonation. For a higher intensity than in our experiment, $I = \unit[4\times 10^{14}]{W/cm^2}$, electron recollision was recently shown to play a significant role for the deprotonation channel, but not for the isomerization channel \cite{Doblhoff-Dier2016}. Sequential double ionization, independent of recollision, indeed leads to a much smaller asymmetry than recollision-induced, non-sequential, double ionization, see e.g. \cite{Kubel2016_pra}. However, the observation of a significant isomerization yield at low intensities is inconsistent with the very low probability of sequential double ionization at low intensities. Due to the low probability of sequential double ionization, recollision probably plays a significant role for the population of all dicationic states, including those leading to isomerization.

\subsection{Isomerization of allene}

\begin{figure}
\centering
\includegraphics[width=\textwidth]{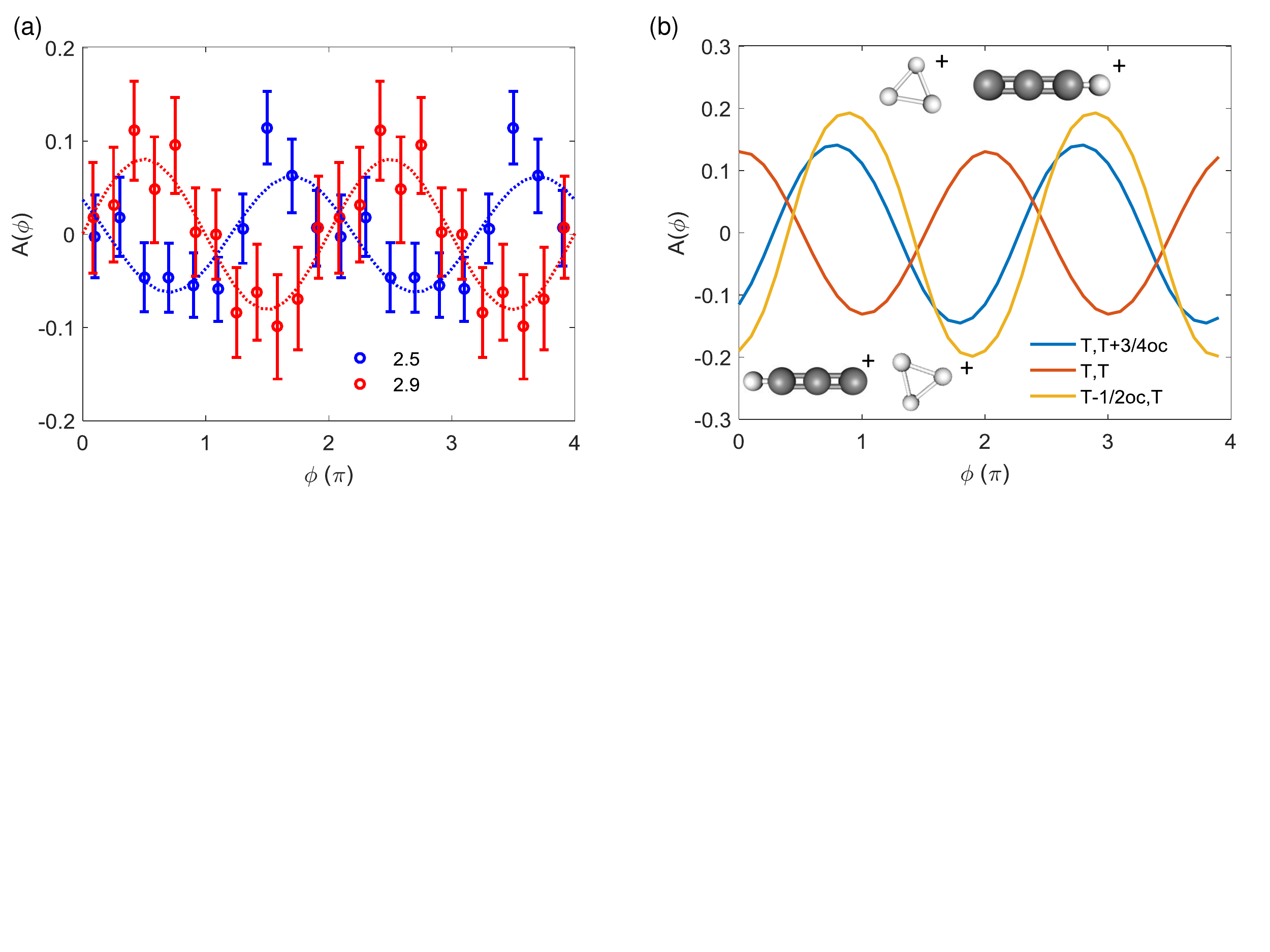}
\vspace{-5.5cm}

\caption{(a) CEP-dependent asymmetries in the momentum distributions of H$_3^+$ ions emitted from allene, for two different intensities, $\unit[2.5 \times 10^{14}]{W/cm^2}$ (blue) and $\unit[2.9 \times 10^{14}]{W/cm^2}$ (red). The data is averaged over a range of $\pm \unit[0.15 \times 10^{14}]{W/cm^2}$ for each intensity. (b) Theoretical predictions for the intensity $\unit[3 \times 10^{14}]{W/cm^2}$ using different double-ionization times. The legend indicates the times $t_1, t_2$ of the first and second ionization, respectively. Here, $T$ denotes the time at which the laser pulse reaches its intensity maximum, oc means optical cycle.} \label{fig:isomerization_allene}
\end{figure}

In allene, hydrogen migration leads to the formation of H$_3^+$, whose emission direction indicates the migration direction. The asymmetry in the momentum distribution of H$_3^+$ ions, detected in coincidence with C$_3$H$^+$, is displayed in Figure \ref{fig:isomerization_allene}a for two intensities. The asymmetry amplitudes recorded for the two intensities ranges are approximately equal. The asymmetry phase, however, differ by nearly $\pi$. The near-complete inversion of the CEP dependence of the isomerization caused by a small intensity change of $\approx 15\%$ is particularly interesting, because it survives the unavoidable averaging over the focal volume, which may wash out features that rapidly vary with changing intensity.

In order to explore whether such a phase shift is compatible with our model of controlling hydrogen migration via vibrational wavepackets, we have performed calculations of the isomerization of allene in which we approximate the effects of different double ionization processes. To this end, we test different ionization times ($t_1, t_2$) at which the first and second ionization steps, respectively, occur. The time delay $t_2 - t_1$ relates to the double ionization mechanism. In sequential double ionization, the second step takes place at the global intensity maximum ($t_2 = T$), and the first ionization is assumed to happen either at the same maximum ($t_1=T$), or a half-cycle earlier ($t_1=T-1/2$\,oc).
For recollision-driven double ionization we assume that the first ionization occurs at the intensity maximum ($t_1 = T$), and the second ionization occurs at $t_2=T + 3/4$\,oc, corresponding to the typical time between emission and recollision of the first liberated electron with the parent molecular ion.

Assuming these three ionization mechanisms, the calculated CEP-dependent asymmetries for the H$_3^+$ emission from allene are displayed in Figure \ref{fig:isomerization_allene}b.
The calculated asymmetry phase for the isomerization of allene exhibits a strong dependence on the ionization times. In particular, the experimentally observed phase shift of nearly $\pi$ can be reproduced, either by the pair $(T,T)$ and $(T-1/2\,\mathrm{oc},T)$, or by $(T,T)$ and $(T, T+3/4\,\mathrm{oc})$. Hence, an alteration in the double ionization dynamics may indeed cause a phase shift in the asymmetry of the isomerization.

\subsection{Double ionization of allene}

A significant alteration in the double ionization dynamics can be expected to leave a signature in the intensity and CEP dependent asymmetry of the center-of-mass motion of the fragments. However, no clear trend can be inferred from the available data on the isomerization fragments. The deprotonation channel of allene, however, provides significantly better statistics, and the center-of-mass momentum distribution of the fragments is well comparable to the one for the isomerization fragments.

\begin{figure}
\centering
\includegraphics[width=\textwidth]{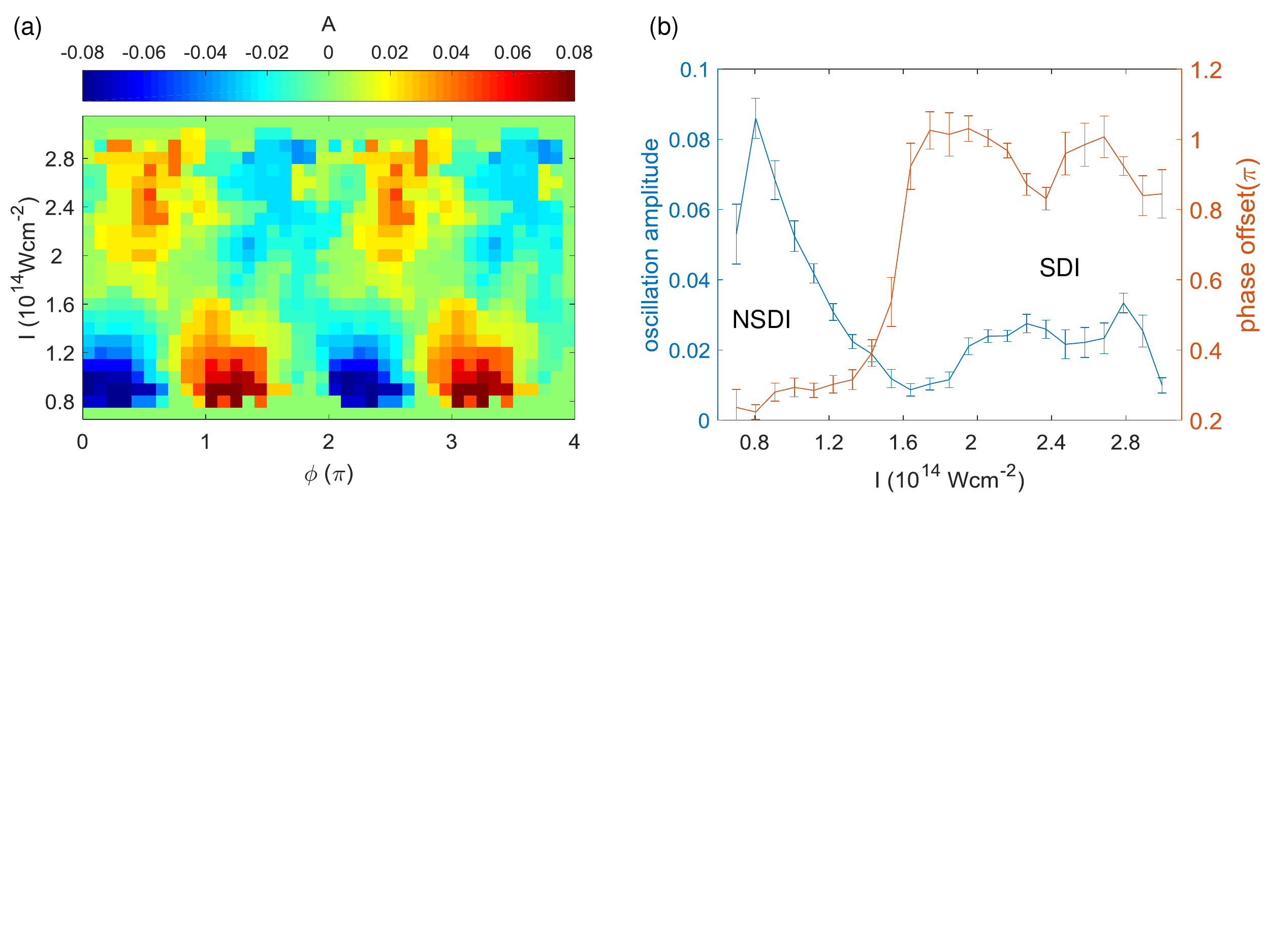}
\vspace{-5.5cm}

\caption{(a) Combined intensity and CEP dependence of the asymmetry in the center-of-mass momentum distributions of the deprotonation fragments of allene (H$^+$ + C$_3$H$_3^{+}$), see text for details. (b) Asymmetry amplitude and phase obtained from fitting (a) with sinusoidals. NSDI and SDI mark the regions of dominant non-sequential and sequential double ioinization.} \label{fig:double_ionization_allene}
\end{figure}

In Figure \ref{fig:double_ionization_allene}, the intensity and CEP-dependence of the asymmetry of the center-of-mass motion of the deprotonation fragments of allene is displayed. The asymmetry amplitude decays with increasing intensity, and almost vanishes at approximately $\unit[1.6 \times 10^{14}]{W/cm^2}$. When the intensity is further increased, the asymmetry amplitude increases again. Coincidentally, the asymmetry phase exhibits a jump of approximately $0.7 \pi$.

The behavior up to an intensity $\unit[1.6 \times 10^{14}]{W/cm^2}$ agrees qualitatively well with the ones observed for acetylene in Figure \ref{fig:double_ionization_acetylene}, and for argon in Ref.~\cite{Kubel2016_pra}. The phase jump and recovery of the asymmetry has not been observed before but agrees with the expectations for sequential double ionization \cite{Kubel2016_pra}. We therefore conclude that the phase jump is consistent with a transition from the non-sequential to the sequential ionization regime.

\section{Conclusion}
We have studied the combined CEP and intensity dependence of ionization and fragmentation processes in the small hydrocarbons acetylene and allene. Our data allows us to characterize the double ionization mechanism, i.e. non-sequential versus sequential double ionization, that initiates deprotonation and isomerization reactions. For the deprotonation of acetylene, we have observed that high intensities are favorable for efficient control using the CEP. This trend agrees with the predictions from a control model in which the CEP dependence of strong-field induced reactions arises from the manipulation of the phases of a vibrational wave packet by the laser. The intensity dependence of the CEP control of the isomerization of acetylene is not reproduced by the theoretical model, which might originate from the rough approximation of the underlying ionization dynamics. For the isomerization of allene, we have observed an inversion of the CEP-dependence, induced by a relatively small change of the laser intensity. We find that the sudden phase jump is consistent with a change in the underlying double ionization mechanism, i.e. a transition from non-sequential to sequential double ionization.

\section*{Acknowledgements}
We are grateful to F. Krausz for his support and fruitful discussions. We are grateful for support by the European Union (EU) through the ERC Grant ATTOCO (No. 307203), by the Max Planck Society, and by the DFG via LMUexcellent and the cluster of excellence Munich Centre for Advanced Photonics (MAP). This project has received funding from the EU’s Horizon2020 research and innovation programme under the Marie Sklodowska-Curie Grant Agreement No. 657544. A.S A. acknowledges support from the American University of Sharjah and the Arab Fund for Economic and Social Development (State of Kuwait). I.B.I. was supported by the Chemical Sciences, Geosciences, and Biosciences Division, Office of Basic Energy Sciences, Office of Science, U.S. Department of Energy.

\section*{Disclosure statement}

The authors declare no conflict of interest.

\bibliographystyle{tfo}
\bibliography{Kubel_mol_phys_v0}

\end{document}